\documentclass{amsart}
\usepackage{amsmath, enumerate}
\usepackage{algorithm}
\usepackage{algpseudocode}
\usepackage{amsfonts}
\usepackage{amssymb}
\usepackage{amsthm}
\usepackage{color}
\usepackage{caption}
\usepackage{graphicx}
\usepackage{url}

\newcommand{\bp}{\begin{problem}}

\newcommand{\ep}{\end{problem}}

\newcommand{\ba}{\begin{answer}}

\newcommand{\ea}{\end{answer}}

\newcommand{\ben}{\renewcommand{\theenumi}{\alph{enumi}}

\renewcommand{\labelenumi}{(\theenumi)}\begin{enumerate}}

\newcommand{\een}{\end{enumerate}}

\newtheorem{defin}{Definition}[section]

\DeclareCaptionType{result}[Table][List of Results]

\title[Cryptography with  groups]{Cryptography with right-angled Artin groups}
\begin{document}

\author[R. Flores]{Ram\'{o}n Flores}
\address{Ram\'{o}n Flores, Department of Geometry and Topology, University of Seville, Spain}
\email{ramonjflores@us.es}
\author[D. Kahrobaei]{Delaram Kahrobaei}
\address{Delaram Kahrobaei, CUNY Graduate Center, PhD Program in Computer Science and NYCCT, Mathematics Department, City University of New York, New York University, Tandon School of Engineering, Computer Science Department}
\email{dkahrobaei@gc.cuny.edu, dk2572@nyu.edu}

\begin{abstract} In this paper we propose right-angled Artin groups as a platform for secret sharing schemes based on the efficiency (linear time) of the word problem. Inspired by previous work of Grigoriev-Shpilrain in the context of graphs, we define two new problems: Subgroup Isomorphism Problem and Group Homomorphism Problem. Based on them, we also propose two new authentication schemes. For right-angled Artin groups, the Group Homomorphism and Graph Homomorphism problems are equivalent, and the later is known to be NP-complete. In the case of the Subgroup Isomorphism problem, we bring some results due to Bridson who shows there are right-angled Artin groups in which this problem is unsolvable.
\end{abstract}



\maketitle
\tableofcontents
\section{Introduction}

Using algorithmic group theoretic problems in cryptography has been an active area of research since 1999 (see \cite{Myasnikov-book} for a thorough account). The complexity of different algorithmic problems (Conjugacy Problems, Membership Problems, etc.) have made available a lot of families of groups as platform groups for cryptographic protocols, as for example:

\begin{itemize}

\item Braid groups, using the Conjugacy Search Problem \cite{KoLee00}.

\item Polycyclic groups, using the Conjugacy Search Problem \cite{EickKahrobaei04}, \cite{GK2016}.

\item Thompson groups, based on the Decomposition Search Problem \cite{ShpilrainUshakov05}.

\item Hyperbolic groups, using properties of subgroup distortion and the Geodesic Length Problem \cite{CKL}.

\item Free metabelian groups, based on the Subgroup Membership Search Problem \cite{VSGZ}, and in the Endomorphism Search Problem \cite{Habeeb-Kahrobaei-Shpilrain}.

\item Free nilpotent $p$-groups, for a semidirect product public key \cite{KS16}.

\item Linear groups \cite{BFX}.

\item Grigorchuk groups, for cryptographic protocols \cite{Pet03}.

\item Groups of matrices, for a Homomorphic Encryption scheme \cite{GP04}.

\end{itemize}

Note that some are infinite and some are finite, but they are all non-commutative.

As mentioned above, we propose here right-angled Artin groups for two secret sharing schemes, as well as two authentication schemes. For the first, we follow the approach of Habeeb-Kahrobaei-Shpilrain \cite{Habeeb-Kahrobaei-Shpilrain} and Shamir \cite{Shamir}, while the second is a modification of two protocols developed by Grigoriev-Shpilrain \cite{Grigoriev-Shpilrain} in the context of graphs, that we adapt to the language of right-angled Artin groups by using the graph which is always associated to these groups.  Then, we take advantage of the fact that many graph-theoretic problems that are proved to be NP-complete can be translated to a group-theoretic setting in the right-angled Artin groups, and also of some unsolvability results which are proper from this context. Besides that it is always of interest to introduce new applications of Group Theory in cryptography, we also note that working with group presentation is easier and sometimes more practical than working with graphs. We note that Shpilrain-Zapata have proposed a key exchange based on Artin groups but this class is much bigger than right-angled Artin groups \cite{VSGZ}.

This note is structured as follows. In Section \ref{RAAGs} we review the main features of right-angled Artin groups that will be useful for our purposes. Section \ref{Sharing schemes} is devoted to the description of the sharing schemes, while the authentication schemes are treated in Section \ref{Auth}. Finally, in Section \ref{complexity} we deal with study of the security of the proposed protocols.

\section{Right Angled Artin Groups}
\label{RAAGs}
 First we will introduce the main facts concerning right-angled Artin groups, a class probably introduced first in \cite{HR71} by Hauschild and Rautenberg (which called them \emph{semifree} groups) in the seventies. Good surveys about the topic can be found in \cite{korbeda} and \cite{charney}, while a good general introduction for the theory presentations of groups is \cite{MKS}.

\begin{defin} [Right-angled Artin groups] Let $\Gamma$  denote a finite simplicial graph. We will write $V = V (\Gamma)$ for the finite set of vertices and $E(\Gamma) \subset V \times V$ for the set of edges, viewed as unordered pairs of vertices. The requirement that $\Gamma$ be simplicial simply means that the diagonal of $V \times V$ is excluded from the set of edges. The right-angled Artin group on $\Gamma$ is the group
$$A(\Gamma) = \langle V|[v_i, v_j] =1 \text{ whenever } (v_i, v_j) \in E \rangle.$$
In other words, $A(\Gamma)$ is generated by the vertices of $\Gamma$, and the only relations are given by commutation of adjacent vertices.
\end{defin}

Observe that right-angled Artin groups, that are associated to a finite simplicial graph (the \emph{Artin graph}), are always finitely presented. It is clear from the definition that there is a bijective correspondence between isomorphism types of right-angled Artin groups and isomorphism types of finite simplicial graphs, in the sense that two right-angled Artin groups $A(\Gamma)$ and $A(\Lambda)$ are isomorphic if and only if $\Gamma = \Lambda$. Moreover, a map $f:A_1\rightarrow A_2$ of right-angled Artin groups is a homomorphism if and only if it induces a graph homomorphism between the corresponding graphs.



We will be specially interested in the subgroups generated by subsets of the set $S$ of generators. If $T\subset S$ is such a subgroup of a right-angled Artin groups $A$, it is usually denoted by $A_T$ and called a \emph{special subgroup} of $A$. Note that every special subgroup of $A$ gives rise to a subgraph of $\Gamma_A$, but the converse is not true. For example, if we consider the graph with vertices $\{v_0,v_1\}$ and edge $[v_0,v_1]$, which corresponds to the free abelian group in two generators, the 0-dimensional subgraph defined by the two vertices produces the free group in two generators, which is not a subgroup of $\mathbb{Z}^2$. It is easy to see that a subgraph $\Gamma'$ of an Artin graph $\Gamma$ defines a special subgroup of the corresponding right-angled Artin groups if and only if $\Gamma'$ is a \emph{full} subgraph of $\Gamma$.

\begin{defin} A subgraph $\Gamma'$ of a graph $\Gamma$ is \emph{full} if for every pair of vertices $\{v,w\}$ in $\Gamma'$ such that $[v,w]$ is an edge in $\Gamma$, $[v,w]$ is an edge in $\Gamma'$.

\end{defin}

The full subgraphs are also called spanning or induced. This condition is important in order to use these subgroups as a platform for authentication.





\section{Secret sharing threshold schemes}
\label{Sharing schemes}

Habeeb-Kahrobaei-Shpilrain have proposed a cryptosystem based on efficiency of the word problem \cite{Habeeb-Kahrobaei-Shpilrain}, and we intend to use it with right-angled Artin groups. Let us describe the two schemes.

In the first protocol, which is an $(n,n)$-threshold scheme, the dealer distributes a $k$-column
$C=\left(
\begin{smallmatrix} c_{1}\\c_{2}\\ \cdot \\ \cdot \\ \cdot \\ c_{k}
\end{smallmatrix}\right)$  consisting of bits (0's and 1's), among $n$
participants in such a way that the column can be reconstructed only
when all participants combine their information. A set of generators $X=\{x_1,\ldots x_m\}$ is public. Then:

\begin{enumerate}

\item The dealer uses a secure channel to assign to each participant $P_j$ a set of commutators $R_j$ of the generators in $X^{\pm}$. Recall that each right-angled Artin group $G_j=\langle x_1,\ldots x_m| R_j\rangle$ has efficiently solvable word problem (see Section \ref{Wordproblem} below).

\item The secret bit column $C$ is split by the dealer in a mod $2$ sum $\sum_{j=1}^nC_j$ on $n$ bit columns, which are secretly distributed to the $n$ participants.

\item Words $w_{1j},\ldots ,w_{kj}$ in the generators of $X$ are openly distributed by the dealer to the participant $P_j$, for every $1\leq j\leq n$. The words are selected in such a way that $w_{ij}\neq 1$ is $c_i=0$ and $w_{ij}=1$ if $c_i=1$.

\item Each participant $P_j$ check, for each $i$, if the word $w_{ij}=1$ in the right-angled Artin group $G_j$ is trivial or not. Then, each participant can make the column $C=\left(
\begin{smallmatrix} c_{1j}\\c_{2j}\\ \cdot \\ \cdot \\ \cdot \\ c_{kj}
\end{smallmatrix}\right)$ , whose entries are 0's and 1's.

\item The secret can be now reconstructed by forming the vector sum $\sum_{j=1}^nC_j$, again with the sum taken mod 2.

\end{enumerate}


The second protocol is a $(t,n)$-threshold scheme, and a modification of the previous one that takes into account some ideas from \cite{Shamir}, and allows a subset of size $t$ of the total number of participants $n$ to reconstruct all the information. Now the secret is
an element $x\in \mathbb{Z}_{p}$, and the dealer chooses a
polynomial $f$ of degree $t-1$ such that $f(0)=x$. In addition the
dealer determines integers $y_{i}=f(i)(\textrm{mod } p)$ that are
distributed to participants $P_{i}$, $1\leq i \leq n$ (we assume that all integers $x$ and $y_{i}$ can be
written as $k$-bit columns). A set of
group generators $\left\{x_{1}, \ldots, x_{m}\right\}$ is
public.

\begin{enumerate}

\item A set of commutators $R_j$ of the generators in $X$ is secretly distributed to the participants by the dealer. The group $G_j=\langle x_1,\ldots , x_m|R_j$ is a right-angled Artin group, and hence it has an efficiently solvable word problem.

\item Now $k$-columns $b_{j}=\left( \begin{smallmatrix} b_{1j}\\b_{2j}\\ \cdot \\ \cdot \\
\cdot \\ b_{kj} \end{smallmatrix}\right)$, with $1 \leq j \leq n$, are openly distributed by the dealer to each participant, being their entries words in the generators. The words $b_{ij}$ are chosen in such a way that, after replacing them by bits (the bit ``1" if $b_{ij}$ is trivial in the right-angled Artin group $G_j$ and ``0" otherwise), the resulting column represents the integer $y_i$.

\item Now for each word $b_{ij}$, the participant $P_j$ checks if this word is trivial or not in the right-angled group $G_j$, and in this way he/she obtains a binary representation of $y_j$.

\item Finally, every participant has a point $y_i=f(i)$ of the original polynomial, and then every set of $t$ participants is able now to obtain $f$ by polynomial interpolation, and also the secret number $x=f(0)$.

\end{enumerate}

Note that every subset of $t$ participants can recover the secret $x$ by constructing the polynomial $f$ via interpolation, and this scheme can be arranged in
such a way that participants do not have to reveal their individual
shares $y_{i}$ to each  other if they do not want to. More details of these protocols can be found in \cite{Habeeb-Kahrobaei-Shpilrain}.

\section{Authentication Schemes Based on the Group Homomorphism and the Subgroup Isomorphism problems}
\label{Auth}
Grigoriev and Shpilrain have proposed in \cite{Grigoriev-Shpilrain} some authentication protocols using graph homomorphisms problem and subgraph isomorphism problem. In the sequel, we introduce two different protocols, which are based on the group homomorphism and the subgroup isomorphism problems, and that we introduce using right-angled Artin groups as a platform; they are inspired by the work of these authors in the sense that they are originally (but not necessarily) addressed to be used with group graphs as a platform. In this sense, and as we will see below, we would make profit of unsolvability results for groups and also for graphs. We remark that we have developed no analogy of the protocol based in the classical Graph Isomorphism problem, as it has recently been shown by Babai \cite{Babai15} that its complexity is quasi-polynomial.

\subsection{An Authentication scheme using Group Homomorphism problem}
\label{Homo}
Consider two finitely presented groups $G_1=\langle S_1|R_1 \rangle$ and $G_2=\langle S_2|R_2 \rangle$, being $S_i$ generators and $R_i$ relations, $i=1,2$. The Group Homomorphism problem asks if there is a homomorphism $G_1\rightarrow G_2$ that takes generators in $S_1$ to generators in $S_2$.

The authentication protocol is the following:

\begin{enumerate}

\item Alice's public key consists of two finitely presented groups $G_1=\langle S_1|R_1 \rangle$ and $G_2=\langle S_2|R_2 \rangle$. Alice's long term private key is a homomorphism $\alpha$ sending generators in $S_1$ to generators in $S_2$.

\item Alice selects another finitely presented group $G=\langle S|R \rangle$, and a homomorphism $\beta:G\rightarrow G_1$ which sends generators in $S$ to generators in $S_1$. Then she sends $G=\langle S|R \rangle$ to Bob, and keeps the homomorphism $\beta$ to herself.

\item Bob chooses a random bit and sends $c$ to Alice.

\begin{itemize}

\item When $c=0$, Alice sends the homomorphism $\beta$ to Bob, and Bob should check if $\beta(G)=G_1$, and if $\beta$ is a homomorphism that takes generators in $S$ to generators in $S_1$.

\item When $c=1$, Alice sends the composite $\alpha \beta$ to Bob, and Bob checks whether $\alpha \beta (G)=G_2$, and if the composite is a homomorphism that takes generators in $S$ to generators in $S_2$.

\end{itemize}

\end{enumerate}



\subsection{An Authentication Scheme Based on the Subgroup Isomorphism problem}
\label{Sub}
The authentication protocol we propose is as follows:
\begin{enumerate}
\item Alice's public key consists of two isomorphic subgroups of a group $\Gamma$, $G_1$ and $G_2$. Alice's long-term
private key is an isomorphism $\alpha : G_1 \rightarrow G_2$.
\item To begin authentication, Alice selects a group $G$ together with the isomorphism $\beta : G \rightarrow G_1$ and sends the group $G$ (the commitment) to Bob, while keeping $\beta$ to herself.
\item Bob chooses a random bit $c$ and sends it to Alice.
\end{enumerate}
\begin{itemize}
\item If $c = 0$, then Alice sends the isomorphism $\beta$ to Bob, and Bob checks whether $\beta(G) = G_1$ and whether $\beta$ is an isomorphism.
\item If $c = 1$, then Alice sends the composition $\alpha \beta = \beta (\alpha)$ to Bob, and Bob checks whether $\alpha \beta (G) = G_2$ and whether $\alpha \beta$ is an isomorphism.
\end{itemize}

\section{Complexity and security}

\label{complexity}

In this section we state the complexity results that make right-angled Artin groups a good platform for the previous protocols.

\subsection{Word problem}
\label{Wordproblem}

To introduce a family of groups as a platform for the secret sharing scheme described \cite{Habeeb-Kahrobaei-Shpilrain} it is necessary that its word problem can be solved efficiently. In the mentioned paper, for example, the authors apply their cryptosystem for small cancellation groups. In the case of right-angled Artin groups, the easiness of the word problem was first proved in a paper by Liu-Wrathall-Zeger \cite{LWZ} which in a more general framework of free partially commutative monoids, describes an algorithm which is effective in linear polynomial time. More recently, Crisp-Goddelle-Wiest \cite{CGW} have extended this result (with different methods) to some families of subgroups of right-angled Artin groups, as for example braid groups.

\subsection{Security assumption, complexity analysis and platform groups (right-angled Artin groups)}

The security of our proposed authentication schemes relies, for the first scheme, on the difficulty of the Graph Homomorphism problem, and for the second, on some Bridson unsolvability results. Let us analyze in detail both situations. 

\subsubsection{Group Homomorphism Problem and proposed authentication scheme:}
We observe that the problem is equivalent to the Graph Homomorphism problem for graphs, as there is a bijection between right-angled Artin groups and finite simplicial graphs (see Section \ref{RAAGs}), and recall that this problem has been shown to remain NP-complete even when the graph in the right is a triangle \cite{GJ}. Hence, it would be enough here to select two right-angled Artin groups $\Gamma_1$ and $\Gamma_2$ such that $\Gamma_2$ contains a free abelian group in three generators.

\subsubsection{Subgroup Isomorphism Problem and proposed authentication scheme:}
Martin Bridson has proved \cite{Br14} that there exist families of right-angled Artin groups for which this problem is unsolvable, even for finitely presented subgroups. Let us briefly recall the construction. He starts with a free group in a finite number of generators, and performs over it Rips construction \cite{Rips82} in the specific version of Haglund and Wise (\cite{HaWi08}, Section 10). In this way we obtain an explicit presentation of a hyperbolic group $\Gamma$ that possess a finite index subgroup $\Gamma_0<\Gamma$, which is the fundamental group of a special cube complex. This complex is subject to certain restrictions (\cite{HaWi08}, Theorem 1.1), that give rise to the existence of a local isometry with a standard cube complex, and in particular imply the existence of an embedding of $\Gamma_0$ in the fundamental group of the latter, which is a right-angled Artin group and we call $A$. The group $\Gamma_0$ also projects onto a non-abelian free group, and the kernel is infinite and finitely-generated. Then, by a previous result of Bridson-Miller \cite{BrMi03}, the subgroup Isomorphism problem is unsolvable for every product $\Gamma_0\times \Gamma_0\times F$ , being $F$ any non-abelian free group. As $\Gamma_0<A$, the problem is also unsolvable for $A\times A\times F$, and this is a right-angled Artin group itself, as it is the product of right-angled Artin groups.

In general, to compare the Subgroup Isomorphism problem and the Subgraph Isomorphism problem we need that the generators and relators on the groups can be represented as a graph. But this is only a necessary condition. For example, in right-angled Artin groups there are plenty of subgroups that cannot be represented by a subgraph of the Artin graph (for example, the cyclic group generated by the product of two generators). An authentication scheme based in this problem for right-angled Artin groups only should make use of the special subgroups, and should take into account the fact that not every subgraph of the Artin graph represents a special subgroup. This approach is closer to the problem of subgroup isomorphism for full subgraphs of a finite graph, usually called the \emph{induced Subgraph Isomorphism problem}, which is known to be NP-complete in general (see \cite{KOSU} for a reference). For the classical Subgroup Isomorphism problem, it is more straightforward to appeal to Bridson unsolvability results described above.

\section*{Acknowledgements}
We thank the referees for their comments and suggestions, which have improved the quality and readiness of this paper.

Delaram Kahrobaei is partially supported by a PSC-CUNY grant from the CUNY Research Foundation, the City Tech Foundation, and ONR (Office of Naval Research) grant N00014-15-1-2164. Part of the work was done while visiting the UPV/EHU in Bilbao funded by the ERC grant PCG-336983, especially we thanks Montse Casals for the fruitful discussions. Delaram Kahrobaei has also partially supported by an NSF travel grant CCF-1564968 to IHP in Paris. Ram\'{o}n Flores is partially supported by MEC grant MTM2010-20692.

\bibliographystyle{plain}

\bibliography{bib}
\end{document}